\RequirePackage{ifpdf}
\pdfoutput=1
\documentclass{JINST}
\usepackage{graphicx}

\title{A light readout system for gas TPCs}

\author{G. Giroux$^{  }$~\thanks{Corresponding author}~\footnote{Present address: Department of Physics, Queen's University,
Kingston, ON, Canada}, M. Auger$ $\footnote{Present address: Physik Institut der Universit\"{a}t Z\"{u}rich, Z\"{u}rich, Switzerland}, D. Franco$ $\footnote{Present address: IPNL, Universit\'{e} Claude Bernard Lyon 1, CNRS/IN2P3,
F-69622 Villeurbanne, France}, M. Weber$ $\footnote{Present address: Physics Department, Stanford University,
Stanford CA, USA}, S. Delaquis$ $, R. Gornea$ $, P. Lutz$ $, J.-L. Vuilleumier$ $  and J.-M. Vuilleumier$ $\\ 
\llap{$  $}Albert Einstein Center for Fundamental Physics, Laboratory for High Energy Physics\\
 University of Berne, Sidlerstrasse 5, CH-3012 Berne, Switzerland\\

E-mail:\email{guillaume.giroux@lhep.unibe.ch}}
\maketitle
\abstract{A novel light detection scheme has been tested for use in medium-pressure
gas TPCs, in view of rare events searches in low energy particle physics. It has 
the advantage of minimal interference with the ionization collection
system, used for track imaging. It provides an absolute time reference, which allows an absolute 
determination of the Z coordinate of events, along the direction of the drift field. This makes possible 
a fiducial cut along the Z-axis, allowing to reduce the background from the ends of the drift volume. }


\begin{document}

\section{Introduction}

Time projection chambers (TPCs) have shown to be powerful tools in low energy particle physics. Gas TPCs at moderate
pressure allow to visualize precisely the particle tracks produced in interactions and decays, recording the spatial development of the ionization signal. 
This makes a stringent selection of events possible, with a corresponding background suppression. This was 
demonstrated in the search for neutrinoless double beta decay in $^{136}$Xe in the Gotthard underground laboratory \cite{Got98}. This feature will be
exploited as well in the NEXT experiment \cite{NEXT}, searching for the same decay. The study of 
$\overline{\nu}_e e^-$ scattering at the Bugey reactor \cite{MUNU05} also took advantage of that. 

The X and Y spatial information, in the plane perpendicular to the drift field, is provided by a position sensitive
readout plane at one end, acting as anode, opposite to the HV cathode. The Z coordinate, parallel to the drift field, is reconstructed from the drift time 
of the ionization charge to the anode. One problem, in non accelerator experiments such as those listed above, is the lack of absolute time 
reference for the events, so that no absolute Z coordinate reconstruction along the drift direction is possible. This is
problematic, as most background sources are located in the materials surrounding the fiducial volume. The cathode is in general quite contaminated, if traces of 
radon, in particular $^{222}$Rn, are present in the gas of the TPC. Positive ions produced in the decay of $^{222}$Rn and its daughters, 
starting with $^{218}$Po, can drift to the cathode where they stick. Without absolute Z coordinate, it is impossible to establish a 
fiducial cut eliminating events from the cathode. To a large extent, the same is true of the anode, a potential background source for similar reasons. 

Moreover an absolute Z determination may be useful in a large size TPC to 
estimate possible charge losses along the drift path, and calculate a 
correction, in order to improve the energy resolution.

Depending on the gas filling the TPC, however, it is possible to observe the 
primary prompt scintillation light accompanying the production of ionization 
charge. 
This can provide a time reference, leading to an absolute Z coordinate 
of events. Achieving that is not completely obvious, as gas amplification is necessary to 
observe tracks from minimum ionizing particles, which requires the addition of a 
quench gas in many interesting cases. The quench gas may absorb the scintillation 
light. A careful  choice is thus necessary. 

A second problem is the position of the light sensors, whether classical photomultipliers, Avalanche Photodiodes (APD) or Silicon Photomultipliers (SiPM). 
One classical solution is to place them behind the cathode, or anode, made of wire grids with reasonable light transmission. 
This geometry was adopted in many TPCs using liquid argon or xenon. 
But this imposes constraints on the anode and cathode construction, which can be problematic. For instance readout planes with 
micropatterns such as Micromegas \cite{Micromegas}, which offer great flexibility, cannot be used. Also the solid angle is 
restricted. Placing the light sensors directly on the side is difficult because of the high electric field gradient. 

In this paper we report on tests of a scheme allowing to detect the light on the side of the TPC, with little interference with the 
ionization collection system. It uses light guides on the sides, coated with a wave length shifter absorbing the primary scintillation 
light, and converting it to longer wavelengths  \cite{TPC_08}. A large fraction of the light emitted in the solid angle for total 
reflection reaches the ends. One end is instrumented with SiPM light sensors, in a region with no or little electric field. 
A diffuse white reflector at the other end, with a gap, reflects back part of the light, which is retrieved.  

We concentrated our tests on two gases suited  for use in TPCs. The first is xenon with 1-2~\% CF$_{4}$. This choice was motivated by 
a possible future study of neutrinoless double beta decay in $^{136}$Xe. The addition of CF$_4$ was shown to give good ionization 
amplification with Micromegas readout planes \cite{Leila}, to suppress charge diffusion, and to increase the ionization drift velocity. 
CF$_4$ is expected to be transparent to the primary VUV scintillation light of xenon centered around 175~nm \cite{ArXe_scint}. 
The second gas which was investigated was pure CF$_4$, which is known to scintillate in a wide range of wavelengths, with a 
broad peak in the UV, and one in the visible. CF$_4$ has a relatively low Z, with corresponding low multiple scattering leading to straighter tracks, which 
can be an advantage \cite{MUNU05,sMUNU}. For comparison we also studied pure argon, which scintillates in the VUV, with a maximal emission around 128~nm, lower than xenon. The measurements were made at a pressure of 300 kPa, at 
which particle tracks, even at sub-MeV energy, develop sufficiently to be reconstructed. 

\begin{figure}[htb]
\begin{center}
\includegraphics[height=10cm]{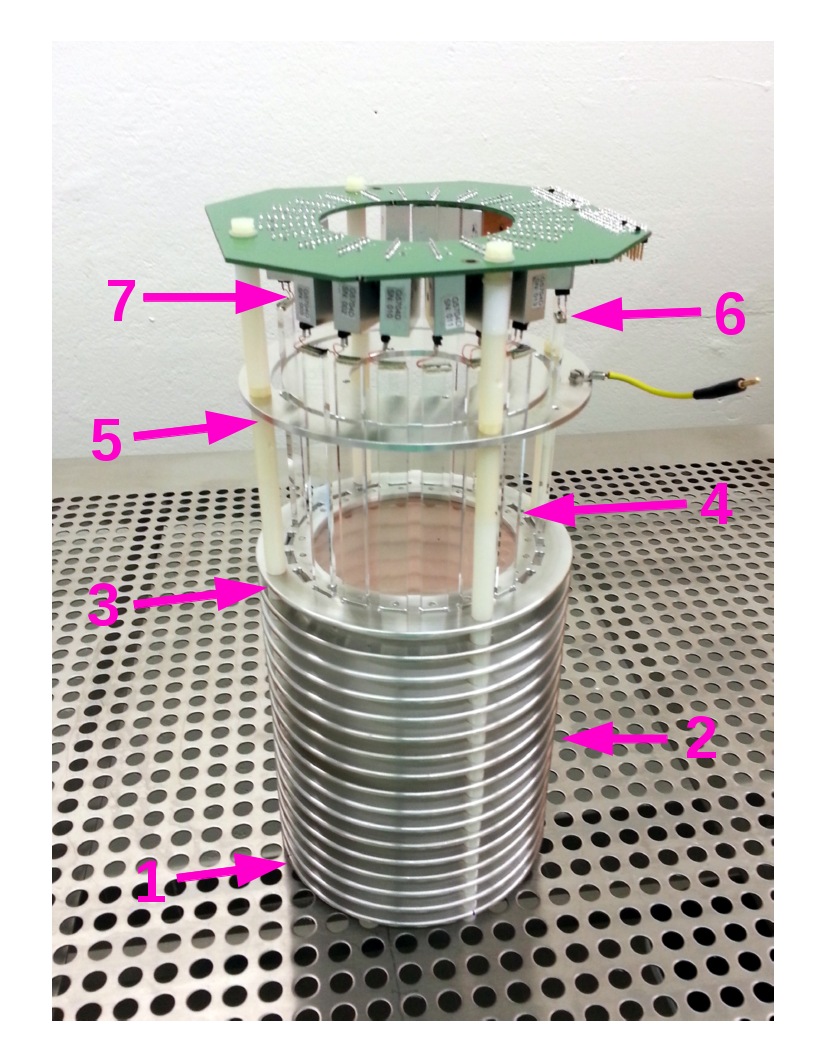}
\includegraphics[height=7cm]{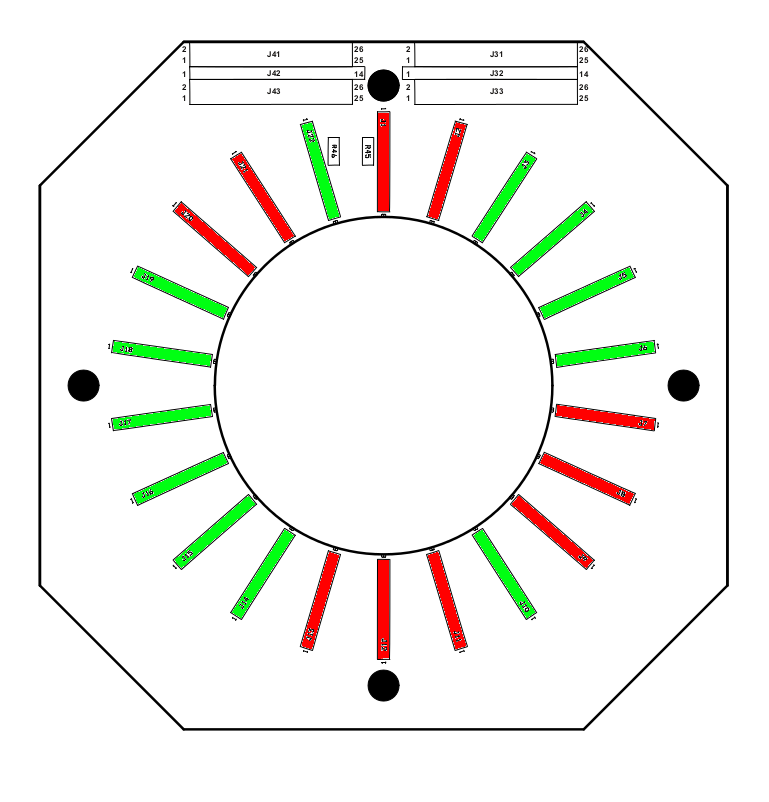}
\caption{Left: the TPC, with (1) the location of the Micromegas plane (not shown), the field shaping rings (2), the cathode (3), the light guides (4), the grounding disc (5), the SiPMs (6) and their preamplifiers and power supplies (7). Right: the 12 
light channels, in green, which were instrumented during the measurements of cosmic muons.}
\label{fi:field_cage}
\end{center}
\end{figure}

\section{Experimental set-up}
The tests were performed in a small TPC with a fiducial volume of 10 cm 
diameter, and 20 cm long, housed in a steel vessel. The vessel can be 
pressurized up to 500 kPa. The fiducial volume is delimited by a field cage made from a Micromegas ionization readout 
plane acting as anode at one end and a cathode made from a mesh on a frame at the other one. Flat field shaping rings between 
the cathode and the Micromegas plane guarantee the field uniformity. 
Rectangular holes are machined in the cathode frame, through which the light 
guides are inserted. They are located inside the field shaping rings. In total 22 light guides can be deployed, although the 
tests discussed in this paper were done with 12 guides only for practical 
reasons. The cage is in the
vertical position, with the readout plane at the bottom, and the cathode at the 
top, as shown in fig. \ref{fi:field_cage}. The gas is constantly circulated 
through an Oxysorb$^{\mbox{\textregistered}}$ filter and a cold trap in the form of a coil 
made from copper tubing immersed in a cooling agent maintained at -70~$^\circ$C. 

The ionization amplification and readout plane of the TPC is of the bulk-Micromegas type \cite{Alexo_mumeg}. The gap between 
the mesh and the resistive layer is 128~$\mu$m. The readout plane contains pixels, which are connected to form orthogonal X and Y 
readout strips read out separately. The pitch is 3.1~mm and the active diameter 10 ~cm, giving 31$\times$31 strips. These 
provide the X and Y information. The Micromegas plane is connected by custom-made flat cables to transimpedance 
preamplifiers located outside the vessel.

\subsection{Light guides}
The dimensions of the light guides are $10\times 3\times 260$~mm. Two materials were considered for the light guides, polystyrene and UV transmitting acrylic. 
Both exhibit an acceptable transmission for the UV light, down to 200~nm for the former \cite{polys_prop}, 300~nm for the latter \cite{acryl_prop}. 
A wavelength shifter from UV to visible Tetraphenyl-Butadiene (TPB), which was shown to have a good efficiency, was chosen \cite{TPB_1,TPB_2}. 
The emission spectrum peaks at 450~nm.

At the emission peak of xenon, and argon, the transmission of either acrylic and polystyrene is limited. For this reason the TPB was deposited 
in thin layers on the surface. The technique described in ref. \cite{TPB_3} was adopted. A mixture of TPB and polystyrene pellets in the 
proportion of 1:3 were dissolved in toluene (50 ml of toluene per gram of polystyrene). The light guides, acrylic and polystyrene, 
were dipped for a few seconds, and then allowed to dry. The polystyrene guides were found to be somewhat hazy near the surface in 
spots. In further tests, when exposed to vacuum, TPB crystallization was observed. For that reason polystyrene was abandoned, 
and all subsequent work was done with the acrylic-based light guides. These are very clear, have a nice finish, and stand vacuum. 
The surface TPB doped film is estimated to be less than 10~$\mu$m thick, which corresponds to a TPB density of less than 2.6$\times$10$^{-4}$ g/cm$^{2}$.

\subsection{Light sensors}

\begin{figure}[htb]
\begin{center}
\includegraphics[height=6cm]{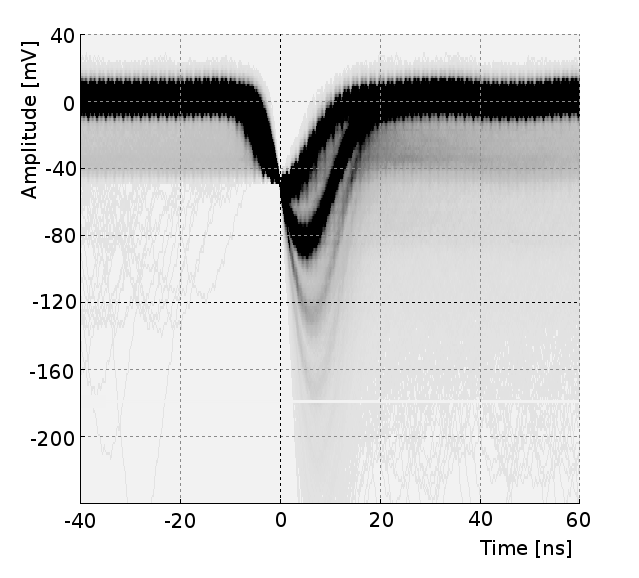}
\includegraphics[height=6cm]{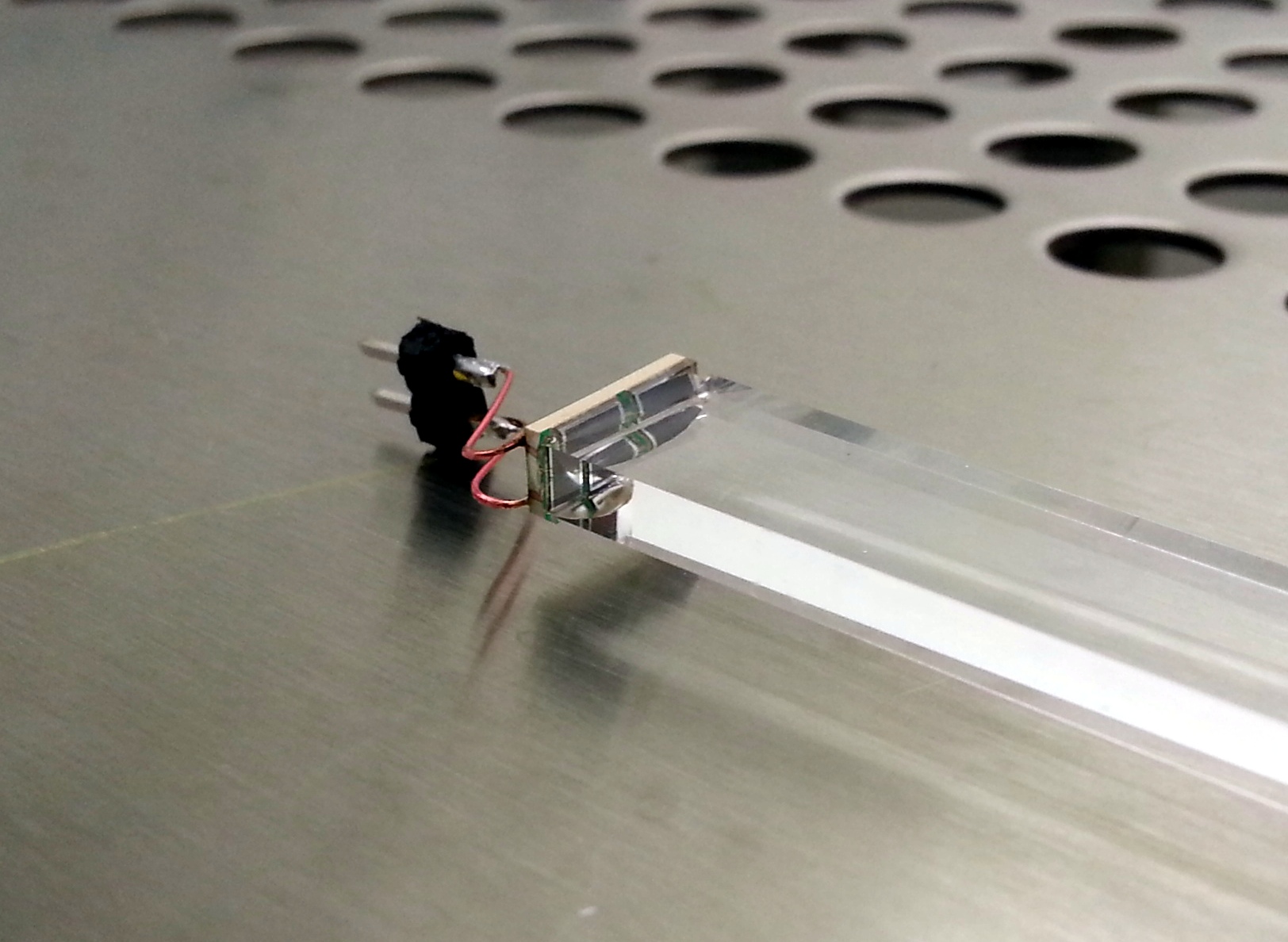}
\caption{Left: Dark counts in a SiPM, the 1, 2, 3 and 4 photoelectron pulses are easily visible; right: the end of a light guide, coated with TPB, with two SiPM's glued on.}
\label{fi:SiPM}
\end{center}
\end{figure}

We used SiPM light sensors made by Hamamatsu, with a sensitive area of 3$\times$3~mm.
The models S10931-050~P (3600 50~$\mu$m$\times$50$~\mu$m pixels, fill factor 61~\%)  and S10931-100~P (900 100~$\mu$m$\times$ 100~$~\mu$m pixels,  fill factor 78~\%) were used. Early tests were also carried out with model S10362-33-050~C, similar to the S10931-050 P, but with a ceramic casing. The 900 pixel version was finally preferred, because of the better fill factor, which leads to a maximal photon detection 
efficiency of 73~\% at 440~nm. It drops to 48~\% for the 3600 pixel SiPM's. In any case the sensitivity well matches the emission spectrum of TPB.
These SiPM's can easily resolve 1, 2, 3 and 4 photoelectron pulses, as shown in fig. \ref{fi:SiPM}. This makes calibration fairly easy.

The SiPM's are glued to the end of a light guide with optical cement. Only one ceramic casing SiPM can be mounted on a light guide. With the other types there is room for two (fig. \ref{fi:SiPM}). The other end of the light guide rests against the support of the Micromegas readout plane, made from Teflon, and acting as diffuse reflector. The SiPM's are placed well behind the cathode of the TPC (fig. \ref{fi:field_cage}). To minimize the electric field around them, a disc with a 5 cm diameter central hole, at ground potential, was placed in front of them.

The pair of SiPM's of each light guide is fed by the same power supply subunit. The two outputs are connected in parallel to a single high speed
transimpedance preamplifier. The power supply regulation and the preamplifiers for all the SiPM pairs are mounted on a circuit board placed behind 
the SiPM's. The voltage applied to each SiPM pair can be adjusted from outside, in the range 72$\pm$10~V, to match the gains, 
nominally at $2.4\times10^{6}$. To reduce the dark current, the preamplifiers can be cooled. The case around them is connected, 
via copper bands, to a tube in which a cooling agent consisting of a mixture of glycol and water circulates. 
The temperature is monitored with PT100 probes. The temperature of the cooling agent is adjusted so that the SiPM's are maintained at 15~$^{\circ}$C.

\section{Tests and characterization of the light detectors}

The light detectors, each composed of a light guide and its light sensors, were tested in various configurations. First measurements were 
performed with  a radioactive $\alpha$ source and with through-going cosmic muons to characterize the light detectors. During these, no voltages were applied 
to the TPC cathode or to the Micromegas amplification system; there was thus no drift field. Finally a drift field was applied, and the 
Micromegas readout plane was used to simultaneously measure the ionization of the events. 

\subsection{Tests with an $\alpha$ source} 

The first tests were performed with two guides, one instrumented with one S10362-33-050~C SiPM, the other one with two S10931-050~P SiPM's, and inserted in
adjacent positions as described above. An $^{241}$Am 5.6 MeV $\alpha$ source with an activity of 37 kBq was placed inside the TPC vessel, on the symmetry axis of the field cage. It could be
moved vertically using a remotely controlled mechanism. In the upper position, at 300 kPa of xenon, the $\alpha$ tracks were contained in 
a volume not seen by the light guides. In the lower position, the source was between the grounding disc and the cathode. The light guides 
were outside of the range of the $\alpha$'s to avoid direct hits.

The singles rate in the SiPM's is fairly high due to dark counts. At an arbitrary pulse height threshold, the coincidence rate between the two light detectors was measured to be 0.2 s$^{-1}$ with
the source in the upper position. It rose to 72 s$^{-1}$ with the source in the lower position. This demonstrates that the 
light guides convert a good fraction of the VUV scintillation light of the xenon to visible light, which is then detected by 
the light sensors. No difference was observed between measurements performed with pure xenon, and a mix of 98 \% xenon and 2 \% CF$_4$. 
This clearly confirms that CF$_4$ does not absorb the VUV light from xenon. 

Tests were also conducted with pure argon and pure CF$_4$ at 300 kPa. With these gases however, part of the $\alpha$ tracks end in the volume seen by the light guides, 
even with the source in the upper position. This reflected in a higher coincidence rate (about 20 s$^{-1}$ in CF$_4$). Nevertheless, with the source in the lower 
position, a significant coincidence rate increase was observed (50 s$^{-1}$ in CF$_4$). This demonstrates that our light detectors can be used in these gases as well.

\subsection{Tests with cosmic muons}
\label{sec:cosmimu}
To be more quantitative and to measure at lower energies, we studied the response of our TPC to cosmic muons. For practical reasons, 12 light detectors only were used for these tests, leaving 10 stations empty, as depicted in fig. \ref{fi:field_cage}. The guides were instrumented with two 10931-100 P SiPM each. 
To minimize the light losses, a Teflon reflector was added between the 
field shaping rings and the light guides. A 1 cm gap was left between its lower edge and the Micromegas readout plane to allow for gas circulation. Four groups of 3 adjacent light detectors were fanned-in together, and their signals sent to a 12-channel charge 
sensitive ADC. The sum signal of the 4 groups was also produced, for visualization on an oscilloscope. 
\begin{figure}[htb]
\begin{center}
\includegraphics[height=7cm]{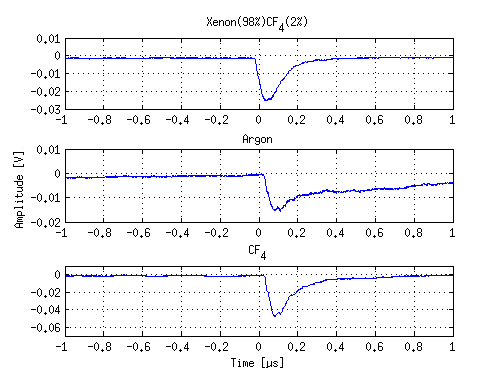}
\caption{Average sum signal of all 12 light detectors from vertical muons with different gases at 300 kPa.}
\label{fi:light_resp}
\end{center}
\end{figure}
\begin{figure}[htb]
\begin{center}
\includegraphics[height=6cm]{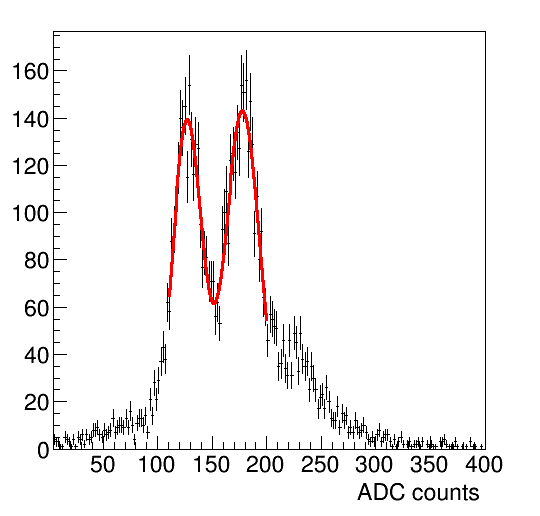}
\includegraphics[height=6cm]{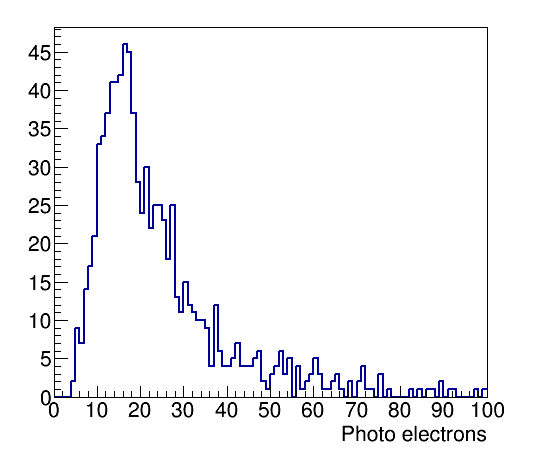}
\caption{Left: 2 and 3 photoelectron peaks observed with the ADC, the threshold was first set at 1.5 photolectron, and then raised to 2.5, to have comparable peak heights; right: Energy deposited by vertical muons crossing the TPC filled with 300 kPa of xenon with 2 \% CF$_4$.}
\label{fi:Xe_muons}
\end{center}
\end{figure}

In a first step nearly vertical muons were 
selected, with two external scintillators above and below the TPC.  The coincidence signal of the external scintillators was used as trigger and time reference. 
The light produced in the fiducial volume of the TPC 
(20 cm long) and in the gap between the 
cathode and the grounding disc (5 cm) has a good probability of reaching one of the light guides. The longest possible 
path of muons in the active volume for light 
is thus 25 cm. On average this reduces however to about 20 cm, corresponding to an energy deposit of 460 keV 
for minimum ionizing muons in 300 kPa of xenon. 

Coincidences between the external scintillators and the light detectors were clearly seen. To get a first idea of the response, the averaged signal of the sum of 
all 4 groups, or 12 light detectors, was produced on an oscilloscope using the external scintillators as trigger. The result is displayed in fig. \ref{fi:light_resp}.
Xenon with 2 \% CF$_4$ is seen to give a very fast signal, with a rise time of order 20~ns. With argon a fast component is seen, 
accompanied by a slower one, with a decay of the order of 1 $\mu$s.
The CF$_4$ signal is somewhat slower than that of xenon. The total light collected, taking into account the slow argon component, is comparable in all three gases.

To be more quantitative the signals of the 4 groups were sent to a 12-channel charge sensitive ADC. The ADC 
was calibrated using the 2 and 3 photoelectron peaks,
visible in fig. \ref{fi:Xe_muons}. For the study of cosmic muons, the ADC was
gated by the external scintillators. Events in which all 4 groups had at least one photoelectron were kept. The sum spectrum of all 4 groups for 300 kPa of xenon with 2 \% CF$_4$ is shown in 
fig. \ref{fi:Xe_muons}. A clear muon peak is seen. Relaxing the cut on the
number of groups with at least 1 photoelectron adds events at low energy,
but does not change the position of the peak. The maximum, corresponding
to minimum ionizing muons, is around 15 photoelectrons. Considering that these muons deposit roughly 460 keV, we conclude that
we observe some  32 photoelectrons per MeV. Knowing that in xenon gas, it takes 72$\pm$6~eV 
to create one primary photon \cite{NEXT_1}\footnote{This value of W$_{s}$ = 72$\pm$6 eV taken from~\cite{NEXT_1} at Xe pressures between 100 and 300 kPa differs significantly from W$_{s}$ = 111$\pm$16 eV~\cite{doCarmo} (100 kPa) and W$_{s}$ = 23.7$^{+7}_{-6}$ eV~\cite{Resnati} (4000 kPa). It is worth noting that these measurements were performed using different radiation sources and at different Xe pressures.}, our system is seen 
to have an overall efficiency of roughly 0.5~\%. Taking into account the 73~\% photon 
detection efficiency of the SiPM's, and the 66 \% coverage of the light guide end by 
the SiPM's, this gives a 1.0~\% efficiency for primary photon collection, wavelength 
conversion and secondary photon collection. With a threshold at the 3 photolectron level, assuming Poisson statistics, a 5 photoelectrons pulse 
has a nearly 90~\% probability of being detected. We thus conclude that we
have a good detection  efficiency above 150~keV.

Similarly, the average number of photoelectrons detected for a nearly vertical cosmic muon 
in CF$_4$ at 300 kPa was measured, and found to be of the order of 20.

\subsection{Operation in the TPC}

Finally, keeping the same configuration, the TPC was turned on. Here 
all measurements were done with xenon and 2~\% CF$_4$ at 300 kPa. First
the cathode was raised to -15 kV with the Micromegas grid left at ground
which gives a drift field of 750~V$\cdot$cm$^{-1}$, or 
2.50~V$\cdot$cm$^{-1}\cdot$kPa$^{-1}$.
The measurement of the averaged signal of the 12 light detectors using the external 
scintillators as cosmic muon trigger, presented in section \ref{sec:cosmimu} for no drift field, 
was repeated. It was found that the light signal is not affected by the electric field at this moderate pressure. A decrease in light induced by the electric field, suppressing electron-ion recombination, has been observed at higher pressure (see ref. \cite{NEXT_2}).

Next the 
cathode voltage was set at -15.9 kV, and the Micromegas grid raised to 
-925~V, yielding the same drift field of 2.50~V$\cdot$cm$^{-1}\cdot$kPa$^{-1}$. 
In that configuration, the light 
detectors measure, in addition to the primary scintillation light, the 
copious secondary light produced in the avalanche in the Micromegas 
gap during the charge amplification process. The two types of light 
pulses are separated in time, except for events passing through the 
anode.

\begin{figure}[htb]
\begin{center}
\includegraphics[height=6cm]{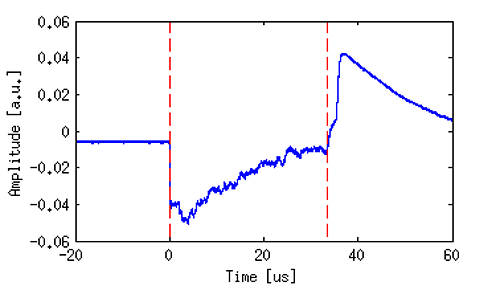}
\caption{Averaged light pulse, dominated by the secondary avalanche 
light, due to nearly vertical muons in the TPC filled with xenon with 
2~\% CF$_4$ at 300 kPa. The differentiation of the preamplifier was 
not corrected for. The vertical dotted red lines indicate the start 
and the end of the pulse. The 33.4~$\mu$s time difference corresponds 
to the total drift time across the 20~cm long drift volume.}
\label{fi:muon_ave}
\end{center}
\end{figure}

First we again investigated nearly vertically through going muons, tagged by 
the external scintillators as in the preceding section. The light 
pulses are nearly rectangular, with a length corresponding to the 
total drift time across the drift volume. The average pulse is shown in fig. 
\ref{fi:muon_ave}. 
The time difference between the leading edge and the trailing edge is 33.4~$\mu$s, corresponding to the total drift time
from cathode to anode, separated by 20~cm. 
This leads to
a drift velocity of 0.60~cm$\cdot \mu$s$^{-1}$, in good agreement with Garfield calculations \cite{Garf}.

Finally, to demonstrate that the primary light pulse can be used to determine a time zero reference, the two external 
scintillators were moved, in order to select muons with a zenith angle around 45$^{\circ}$. A number of these leave the 
fiducial volume of the TPC sideways, before reaching the Micromegas readout plane. 
\begin{figure}[t!]
\begin{center}
\includegraphics[width=10.0cm]{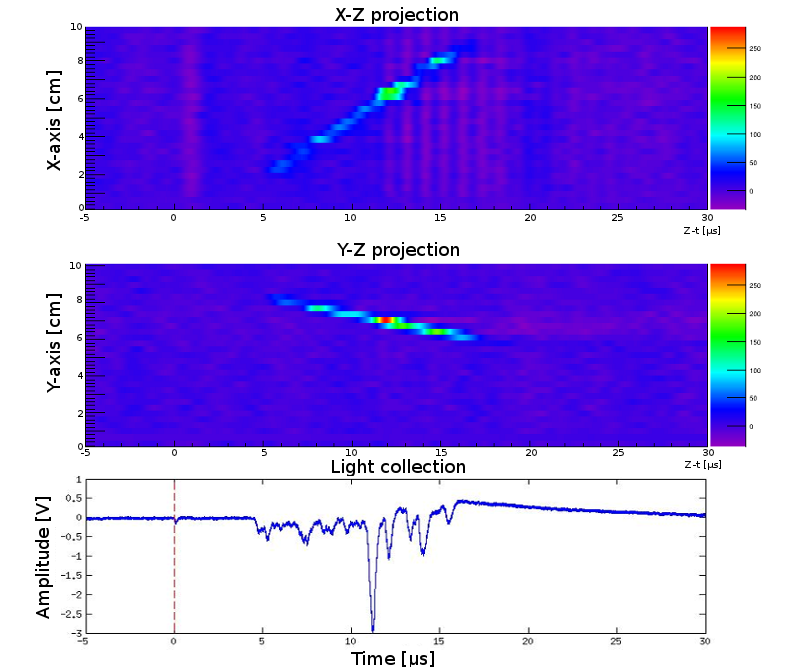}
\includegraphics[width=10.0cm]{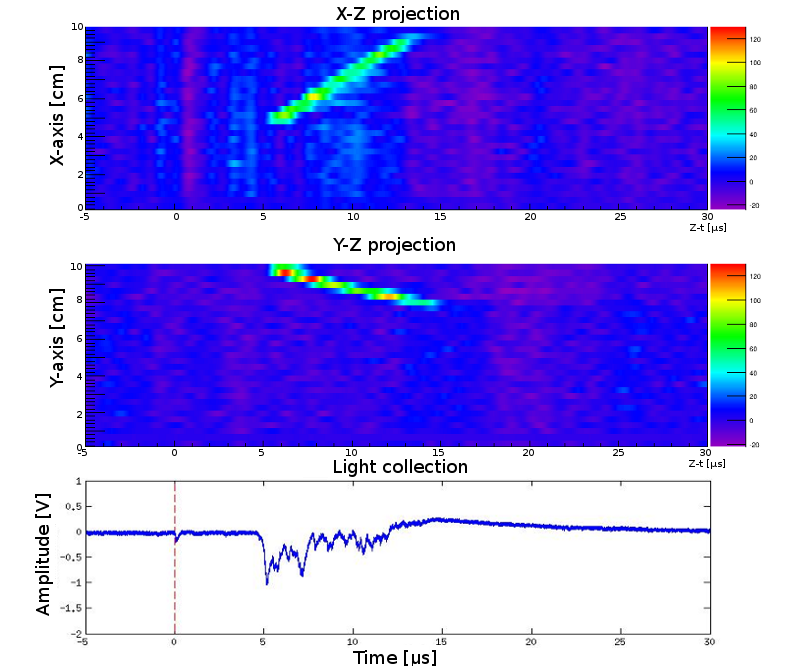}
\caption{Two cosmic muons with a zenith angle around 45$^{\circ}$, selected by the external scintillators, 
which give the time reference; Shown are the time evolution of the light signal (bottom) and the X and Y 
ionization signals from the Micromegas; The primary scintillation light signal in coincidence with the 
external scintillators (dotted vertical red line) at time zero is small but clearly visible, as well as the 
secondary light pulse which follows; the latter is in good correspondence with the X-t(Z) and Y-t(Z) projections 
of the events reconstructed from the ionization signal. In the top event a $\delta$ electron is visible in both 
the light and ionization channels.} 
\label{fi:muon45}
\end{center}
\end{figure}

Two examples are given in fig. \ref{fi:muon45}. The small primary light signal, in coincidence 
with the signal from the external scintillators, is clearly visible. So is the secondary 
light, the time evolution of which follows that of the ionization signals from the Micromegas 
readout plane in both the X-t(Z) and Y-t(Z) projections of the events.
In the top event a $\delta$ electron is visible, in the light channel as well as in the ionization 
channels.

With our coordinate system (Z=0 is on  the anode), a muon enters from the right (in reality top) and travels to the 
left (bottom). The time delay between the prompt scintillation and the leading edge of the secondary
light signal allows to determine the Z coordinate at which the muon left the 
fiducial volume through the side. The precision is given by the rise time of the 
secondary light signal, of order 200~ns, much slower than the primary light signal.

For the top event, the delay is 4.50$\pm$0.20~$\mu$s, corresponding to 
Z=2.70$\pm$0.12~cm. Similarly, using the trailing edge of the secondary light pulse, one sees that the muon 
entered the fiducial volume at Z=9.35$\pm$0.12~cm.

The segment of track has a total length of 9.5~cm, corresponding to an energy deposit of 220 keV, assuming the 
muon to be minimum ionizing, and neglecting the
$\delta$ electron.  

The bottom event shows a muon entering the fiducial volume at Z=8.57$\pm$0.12~cm and leaving it at Z=3.00$\pm$0.12~cm. 
The length of the track segment is 7.5~cm, corresponding to an energy deposit of order 175 keV.

In both cases, the primary light signal is of the order of a few photoelectrons, in agreement with the energy deposited.

\section{Conclusion}

We have demonstrated the feasibility of a light readout system suited for gas TPCs.
It uses light guides mounted on the side of the TPC. They have a thin coating doped 
with TPB, to shift the VUV primary scintillation light, produced by many popular 
gas fills such as argon, xenon and CF$_4$, to visible light, which is in turn 
measured with SiPM's. Good detection efficiency is achieved with our set up down to 150 keV, at which 5 photoelectrons are seen. It should be possible to enhance the signal and lower 
further the detection threshold by improving the coverage. For practical reasons we only mounted 12 
light guides, while 22 could be installed. It might 
be interesting to look for ways to make the TPB coating somewhat thicker, 
to improve the conversion efficiency from VUV to visible. Success is not 
guaranteed however, because of the short absorption length of VUV light in 
polystyrene.

We have also shown that the secondary light produced in the amplification avalanche 
in the ionization readout anode, in our case a Micromegas, is easily detected. The 
time difference between the prompt primary pulse and the leading edge of the secondary pulse
provides a precise absolute determination of the event position in Z, along 
the drift field. This solves a long-lasting problem in the operation of gas TPCs 
in rare events searches at low energy. Here also it may be interesting to 
investigate if the measured secondary light can be used to improve the 
measurement of energy.

Our tests were conducted at 300 kPa, a pressure at which low sub-MeV tracks develop 
sufficiently to do tracking. Working at a higher pressure should be easy with the 
light readout system. The measurement of the ionization would be more problematic. 
We operated the Micromegas at the maximal gain, before discharges occur. The maximal 
achievable gain decreases with pressure. It must be said however that our 
preamplifiers on the ionization channels do not have the lowest possible noise. 
Improvements are possible on that side.

\acknowledgments

The authors wish to thank R. H\"{a}nni for his contribution in building the  
mechanical set-up, C. Tognina for her help in assembling the electronics,
I. Kreslo for useful comments
and R. de Oliveira (CERN) for his advice on Micromegas operations. This work was 
supported by the Swiss National Science Foundation.

\end{document}